\documentclass[journal]{IEEEtran}
\usepackage{amsmath,amssymb}
\usepackage[dvips]{graphicx}
\usepackage{amsfonts}
\usepackage[mathscr]{eucal}
\usepackage{latexsym}
\usepackage{amsthm}
\usepackage{exscale}
\usepackage[mathscr]{eucal}
\usepackage{bm}
\usepackage[dvipsnames]{color}
\usepackage{cases}
\usepackage{epsfig}
\usepackage[center,small]{caption}
\usepackage{algorithm}
\usepackage{algorithmic}
\usepackage[verbose,nospace,sort]{cite}
\usepackage{tabularx}
\usepackage{multirow}
\usepackage{multicol}
\usepackage{balance}
\usepackage{caption}
\usepackage{subcaption}

\scrollmode

\newtheorem{theorem}{Theorem}

\newtheorem{lemma}{Lemma}

\newcommand{\lb}{\mathrm{lb}}
\newcommand{\succf}{\mathrm{succ}}
\newcommand{\thr}{\mathrm{th}}

\graphicspath{{./figsV6/}}

\begin{document}
\title{Trajectory Optimization for Completion Time Minimization in UAV-Enabled Multicasting}
\author{\IEEEauthorblockN{Yong~Zeng,~\IEEEmembership{Member,~IEEE,}   Xiaoli~Xu, and Rui~Zhang,~\IEEEmembership{Fellow,~IEEE}}
\thanks{Y. Zeng and R. Zhang are with the Department of Electrical and Computer Engineering, National University of Singapore, Singapore 117583 (e-mail: \{elezeng, elezhang\}@nus.edu.sg). X. Xu is with the School of Electrical and Electronic
Engineering, Nanyang Technological University, Singapore 639801. (email: xu0002li@e.ntu.edu.sg).}
\thanks{Part of this work has been submitted to the IEEE Wireless Communications and Networking Conference (WCNC), Barcelona, Spain, April 15-18, 2018.}
}

\maketitle

\begin{abstract}
This paper studies an unmanned aerial vehicle (UAV)-enabled multicasting system, where a UAV is dispatched to disseminate a common file to a number of geographically distributed ground terminals (GTs). Our objective is to design the UAV trajectory to minimize its mission completion time, while ensuring that each GT is able to successfully recover the file with a high probability required. We consider the use of practical random linear network coding (RLNC) for UAV multicasting, so that each GT is able to recover the file as long as it receives a sufficiently large number of coded packets. However, the formulated UAV trajectory optimization problem is non-convex and difficult to be directly solved. To tackle this issue, we first derive an analytical lower bound for the success probability of each GT's file recovery. Based on this result, we then reformulate the problem into a more tractable form, where the UAV trajectory only needs to be designed to meet a set of constraints each on the minimum connection time with a GT, during which their distance is below a designed threshold. We show that the optimal UAV trajectory only needs to constitute connected line segments, thus it can be obtained by  determining first the optimal set of waypoints and then UAV speed along the lines connecting the waypoints. We propose practical schemes for the waypoints design based on a novel concept of virtual base station (VBS) placement and by applying convex optimization techniques. Furthermore, for given set of waypoints, we obtain the optimal UAV speed over the resulting path efficiently by solving a linear programming (LP) problem. Numerical results show that the proposed UAV-enabled multicasting with optimized trajectory design achieves significant performance gains as compared to benchmark schemes.
\end{abstract}

\begin{IEEEkeywords}
UAV communication, multicasting, trajectory optimization, network coding.
\end{IEEEkeywords}

\section{Introduction}
Wireless communication systems have gradually evolved to aim not only for high throughput, but also for ultra-reliability, low energy consumption, and supporting highly diversified applications with heterogeneous quality-of-service (QoS) requirements  \cite{613}. To this end, research efforts in the past have mainly focused on conventional networking architectures  typically with fixed infrastructures such as ground base stations (BSs), access points, and relays, which fundamentally limit their capability to meet the increasingly multifarious service requirements cost-effectively. To address this issue, there have been growing interests in providing wireless connectivity from the sky, by utilizing various airborne platforms such as balloons \cite{910}, helikites \cite{912}, and unmanned aerial vehicles (UAVs) \cite{911}, \cite{649}. In particular, wireless communications by leveraging the use of low-altitude UAVs (typically at altitude within one kilometer above the ground) are appealing due to their many advantages, such as the ability of on-demand and swift deployment, high  flexibility with fully-controllable mobility, and high probability of having line-of-sight (LoS) communication links with the ground terminals (GTs) \cite{649}. Therefore, with the continuous cost reduction and endurance improvement of UAVs, together with the device miniaturization of communication equipment, it is anticipated that UAV-enabled communications will play an increasingly more important role in future wireless systems.

Depending on the practical applications, UAVs in wireless communication systems could either be deployed quasi-stationarily at predetermined locations, or fly contiguously  over the served GTs by following certain trajectories. In the former case, one typical application is  UAV-enabled ubiquitous coverage, where UAVs are deployed to assist the existing ground BSs, if any, to ensure seamless wireless coverage for the GTs within a service area \cite{642}, \cite{sharma2016uav}. In this case, the UAVs resemble all essential functionalities of the conventional terrestrial BSs, but typically at a much higher altitude. 
Some practical scenarios for this application include UAV-enabled  offloading in hot spot areas and fast communication service recovery after natural disasters. Along this direction, significant research efforts have been devoted to optimizing the UAV placement in two dimensional (2D) or 3D space \cite{793,803,917,886,926,914}, by exploiting the unique channel characteristics of the UAV-ground links. On the other hand, in the case with flying UAVs for applications such as UAV-enabled mobile relaying \cite{641} and UAV-enabled information dissemination or data collection \cite{918}, the fully controllable mobility of UAVs offers new degrees of freedom in the system design. This can help to significantly enhance the performance compared to conventional systems with fixed relays/BSs on the ground, by dynamically adjusting the UAV positions according to  the locations of the served GTs and their communication requirements \cite{649}. For instance, for UAV-enabled data collection in Internet of Things (IoT) \cite{915} and machine type communications, the UAV can fly close to each of the GTs sequentially so as to shorten their link distance for more energy-efficient data gathering \cite{918}, \cite{887}.  For such applications, the system performance  critically depends on the UAV trajectories, which need to be optimally designed.

Trajectory design or path planning has been a major research area in the existing literature on UAVs. However, prior works mainly focus on UAV navigation applications to ensure its safe fly between a pair of predetermined initial and final locations, under various practical constraints such as collision avoidance with other UAVs and/or terrain obstacles \cite{620,790,920,921}. There have been a handful of works recently on the UAV trajectory design dedicated to optimizing the communication performance. For example, by assuming that the UAV is equipped with multiple antennas and flies with a constant speed, the authors in \cite{659} proposed an algorithm to dynamically adjust the UAV's heading to maximize the ergodic sum rate of the uplink communications from the GTs to the UAV. In \cite{641}, for UAV-enabled mobile relaying systems, a design framework for jointly optimizing the communication power/rate allocation and the UAV trajectory, including both the flying direction and speed, was proposed to maximize the communication throughput. For the non-convex UAV trajectory optimization, \cite{641} proposed the use of successive convex optimization technique to find efficient suboptimal solutions. This technique has been later adopted for UAV trajectory optimization in various other setups, including the energy efficiency maximization for UAV-enabled communication \cite{904}, throughput maximization for UAV-enabled multi-user downlink communication \cite{919}, and sensor energy minimization in UAV-enabled data collection \cite{918}.

\begin{figure}
\centering
\includegraphics[scale=0.4]{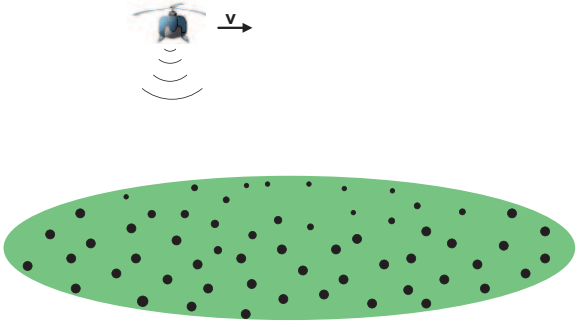}
\caption{UAV-enabled information multicasting.}\label{F:SystemModel}
\end{figure}

In this paper, we study a new UAV-enabled multicasting system as shown in Fig.~\ref{F:SystemModel}, where a UAV is dispatched to disseminate a common file to a set of geographically distributed GTs \cite{930}. UAV-enabled information dissemination or multicasting is one important use case of UAV-enabled communication systems \cite{649},  with a variety of applications such as for public safety and emergency responses \cite{615}, video streaming \cite{924}, \cite{925}, and intelligent transportation systems \cite{923}. Different from the conventional multicasting with static transmitters (e.g., terrestrial BSs), where the multicasting performance is fundamentally limited by the bottleneck link of the user that is most far away from the transmitter, UAV-enabled multicasting is able to overcome this issue by exploiting its high mobility via adaptive trajectory design, which is the main focus of this work.

 Specifically, under a general flat-fading channel model between the UAV and GTs, our objective is to design the UAV trajectory to minimize its mission completion time, while ensuring that each GT is able to recover the file with a success probability no smaller than a given target. Mission completion time minimization is a desirable goal in practice due to the limited UAV on-board energy and hence endurance time. We consider the use of random linear network coding (RLNC) \cite{878} for  UAV multicasting, since it is known to be a robust practical coding technique for such applications with random packet erasures and without the need of dedicated receiver feedback for ARQ (Automatic Repeat reQuest). With RLNC, each GT is able to successfully recover the file as long as it can reliably receive a sufficiently large number of coded packets, whose probability critically depends on the UAV trajectory design. Due to the fundamentally different setups and design objectives, existing UAV trajectory designs (in e.g., \cite{641}, \cite{904}), which are typically for throughput maximization with independent messages for the GTs under a given mission time constraint, are no longer applicable for the new problem considered in this paper, thus calling for new problem formulation and solutions. The main contributions of this paper are summarized as follows.

  First, for UAV-enabled multicasting systems with RLNC, we formulate the optimization problem to minimize the mission completion time, while ensuring that each GT is able to successfully recover the file with a targeting probability, subject to the UAV's maximum speed constraint. The formulated problem is difficult to be directly solved, since the file recovery probability of each GT is a complicated function of the UAV trajectory. To tackle this issue, we derive an analytical lower bound for the file recovery probability of each GT by introducing an auxiliary distance parameter $D$. The main idea is to ignore the portion of the UAV flight time during which the horizontal distance with each GT of interest is greater than $D$,  hence incurring relatively higher packet loss probabilities than a threshold value (specified by $D$). 
    As a result, the UAV trajectory design is reformulated to meet a corresponding constraint on the minimum connection time with each GT, during which their distance is below the critical distance $D$.

 Next, we show that for the reformulated problem, the optimal UAV trajectory only needs to constitute connected line segments. Thus, the problem is further reduced to finding a set of optimal waypoints for the UAV trajectory, and then optimizing the instantaneous UAV speed along the lines connecting these waypoints. However, finding the optimal waypoints is challenging since it is a generalization of the classic Travelling Salesman Problem (TSP) \cite{TSP, 909, 908}, which is known to be NP hard. We thus propose two practical waypoints design schemes based on a novel concept of virtual base station (VBS) placement and by applying convex optimization techniques. Furthermore, for a given waypoint design, we obtain the optimal UAV speed over the resulting path efficiently by solving a linear programming (LP) problem.

  Finally, numerical results are provided to validate the performance of the proposed designs. It is shown that compared to the heuristic benchmark  waypoint designs, 
     the proposed designs can significantly reduce the required mission completion time. Furthermore, as compared to the conventional multicasting setup with a static transmitter, the proposed UAV-enabled multicasting with optimized trajectory achieves significant performance gains in terms of file recovery probability and/or mission completion time. This demonstrates the great potential of UAV-enabled information multicasting in future wireless systems.

The rest of this paper is organized as follows. Section~\ref{sec:SystemModel} presents the system model and problem formulation. In Section~\ref{sec:reFormul}, the lower bound of the file recovery probability is derived, based on which the optimization problem is reformulated. In Section~\ref{sec:PropSol}, the proposed UAV trajectory designs are presented. Section~\ref{sec:numerical} provides the numerical results, and finally we conclude the paper in Section~\ref{sec:Conl}.

{\it Notations:} In this paper, scalars are denoted by italic letters. Boldface lower-case letters denote vectors. $\mathbb{R}^{M\times 1 }$ denotes the space of $M$-dimensional real-valued vectors. For a vector $\mathbf a$, $\|\mathbf a\|$ represents its Euclidean norm. 
 $\log_2(\cdot)$ denotes the logarithm with base $2$. $\mathbb E[\cdot]$ denotes the statistical expectation and $\Pr(\cdot)$ represents the probability. $\mathrm{Bern}(p)$ represents the Bernoulli distribution with success probability $p$, $\mathcal {B}(N,p)$ denotes the binomial distribution with $N$ independent trials each with success probability $p$, and $\mathcal{N}(\mu, v^2)$ denotes the Gaussian distribution with mean $\mu$ and variance $v^2$. For a time-dependent function $\mathbf q(t)$, $\dot{\mathbf q}(t)$ denotes the first-order derivative with respect to time $t$. For a set $\mathcal M$, $|\mathcal M|$ denotes its cardinality. For two sets $\mathcal M_1$ and $\mathcal M_2$, $\mathcal M_1\subset \mathcal M_2$ denotes that $\mathcal M_1$ is a subset of $\mathcal M_2$.

\section{System Model and Problem Formulation}\label{sec:SystemModel}
As shown in Fig.~\ref{F:SystemModel}, we consider a wireless communication system consisting of $K$ GTs denoted by the set $\mathcal K=\{1,\cdots, K\}$, with the location of GT $k$ denoted as $\mathbf w_k\in \mathbb{R}^{2\times 1}$, $k\in \mathcal K$. The GTs' locations are assumed to be known for the UAV trajectory design. A UAV flying at a constant altitude $H$ above the ground is dispatched to disseminate a common information file of total size $W$ bits to all the $K$ GTs. Note that in practice, $H$ could correspond to the minimum altitude to ensure safe UAV operations, e.g., for obstacle avoidance without frequent aircraft ascending or descending.

\subsection{Random Linear Network Coding}
  We assume that RLNC \cite{878} is employed for the UAV transmission, where the information file is linearly coded in the packet level. Specifically, denote the  size  of each packet as $R_p$ bits/pakcet. Then the total number of information packets is $N'= W/R_p$, which are linearly combined with randomly generated coding coefficients from a finite field to obtain $N>N'$ coded packets.\footnote{For convenience, we assume that $W$ is an integer multiple of $R_p$.} These coded packets are then broadcasted by the UAV's transmitter to the GTs along its flight trajectory. 
  As the randomly generated coding coefficients in RLNC are linearly independent almost surely for a sufficiently large field size, each GT will be able to recover the information file as long as {\it any} $N'$ out of the $N$ coded packets are successfully received. Note that for assisting the file recovery based on the network coded packets at the GTs, only the seeds used for generating the random coding coefficients need to be appended with the payload of each packet, and hence the network coding overhead is negligible.

Denote by $R$ the transmission rate in bits/second (bps), which is assumed to be predetermined and remain constant. Then the time required to complete one packet transmission is $T_p=R_p/R$ in second. As a result, the mission completion time, or the total time required to complete the transmission of the $N$ coded packets, is given by
 \begin{align}
 T=NT_p=\frac{W}{R}\frac{N}{N'}. 
 \end{align}

\subsection{Channel Model}
 Within the mission completion time, denote by $\mathbf q(t)\in \mathbb{R}^{2\times 1}$, $0\leq t\leq T$, the UAV's flying trajectory projected onto the ground. Further denote by ${V}_{\max}$ the maximum UAV speed in meter/second (m/s). We then have the constraint $\|\dot{\mathbf q}(t)\|\leq {V}_{\max}$, $\forall t$. The time-dependent distance between the UAV and the GTs can then be written as
\begin{align}
d_k(t)=\sqrt{H^2+\|\mathbf q(t)-\mathbf w_k\|^2}, \ 0\leq t \leq T, \ \forall k\in \mathcal K. \label{eq:dkt}
\end{align}

For a general flat-fading channel model for the UAV-to-GT links, with the $N$ coded packets transmitted by the UAV during the time horizon $T$, the probability that each of the $K$ GTs reliably receives at least $N'$ packets to successfully recover the information file critically depends on the UAV's trajectory $\mathbf q(t)$, $0\leq t \leq T$. 
Our objective in this paper is to optimize $\mathbf q(t)$ so as to minimize the total mission completion time $T$, or equivalently the total number of coded packets $N$ that need to be transmitted, while ensuring that each of the $K$ GTs is able to recover the information file with a success probability no smaller than a given target $\bar{P}$. Note that for practical information multicasting systems, a subsequent device-to-device (D2D) packet sharing phase could be employed, so that those GTs who fail to recover the file will receive additional packets from their peers until they can also successfully recover the file \cite{649}. By increasing the targeting threshold $\bar P$ for the UAV multicasting phase, in general less packets need to be shared in the D2D phase. In this paper, we focus on the UAV multicasting phase, whereas a joint investigation of the UAV multicasting and D2D file sharing would be an interesting problem for future research.

  For the ease of exposition, the time horizon $T$ is discretized into $M$ equally spaced time slots, i.e., $T=M\delta_t$, with $\delta_t$ denoting the elemental slot length, which is appropriately chosen so that the distance between the UAV and the GTs can be assumed to be approximately constant within each slot. For instance, $\delta_t$ might be chosen such that $\delta_t V_{\max} \ll H$. Thus, the UAV trajectory $\mathbf q(t)$ over the time horizon $T$ can be approximated by the $M$-length sequence $\{\mathbf q[m]\}_{m=1}^M$, where $\mathbf q[m]\triangleq \mathbf q(m\delta_t)$ denotes the UAV's horizontal location at time slot $m$.
Furthermore, the UAV speed constraint can be expressed as
\begin{align}
\big \|\mathbf q[m]-\mathbf q[m-1]\big\| \leq \tilde{V}_{\max}\triangleq \delta_t V_{\max}, \ m=2,\cdots, M.
\end{align}
The distance between the UAV and the GTs in \eqref{eq:dkt} can be discretized as
\begin{align}
d_k[m]= \sqrt{H^2+\|\mathbf q[m]-\mathbf w_k\|^2}, \ 1\leq m\leq M, \  k \in \mathcal K.\label{eq:dkm}
\end{align}
The average channel power gain from the UAV to GT $k$ at slot $m$ can be modeled as
\begin{align}
\beta_k[m]=\beta_0d_k^{-\alpha}[m]=\frac{\beta_0}{(H^2+\|\mathbf q[m]-\mathbf w_k\|^2)^{\alpha/2}},\label{eq:betakm}
\end{align}
where $\beta_0$ denotes the channel power gain at the reference distance of $d_0=1$m, and $\alpha\geq 2$ is the path loss exponent.

 With the slot duration fixed to $\delta_t$, the number of packets that can be transmitted by the UAV during each time slot is $L=\delta_t/T_p = R\delta_t/R_p$. 
 For convenience, we assume that $L\geq 1$ is an integer. It then follows that the total number of transmitted packets by the UAV is related to $M$ as $N=ML$. The relationship between the different parameters of the considered system is summarized in Table~\ref{table:basicSetup}.

\begin{table}
\centering
\caption{List of parameters.}\label{table:basicSetup}
\begin{tabular}{p{4cm}|p{3cm}}
\hline
Information file size  & $W$  bits\\
\hline
Packet size & $R_p$ bits \\
\hline
Number of information packets & $N'=W/R_p$ \\
\hline
Number of network coded packets & $N>N'$ \\
\hline
UAV transmission rate & $R$ bits/second \\
\hline
Time for transmitting one packet & $T_p=R_p/R$ seconds \\
\hline
Mission completion time & $T=NT_p=\frac{W}{R}\frac{N}{N'}$ seconds\\
\hline
Time slot length & $\delta_t$ seconds \\
\hline
Number of time slots & $M=T/\delta_t$ \\
\hline
Number of transmitted packets per slot & $L=\delta_t /T_p=N/M$\\
\hline
\end{tabular}
\end{table}

 We assume quasi-static fading channels, where the instantaneous channel coefficients between the UAV and GTs remain unchanged for each packet duration of $T_p$ seconds, and may vary across different packets. Therefore, the instantaneous channel gains between the UAV and GT $k$ can be modeled as
 \begin{align}
 h_k[m,l]=\sqrt{\beta_k[m]} g_k[m,l], \ m=1,\cdots, M, \ l=1,\cdots, L,\label{eq:hkml}
 \end{align}
 where $h_k[m,l]$ denotes the channel coefficient between the UAV and GT $k$ during the transmission of the $l$th packet in time slot $m$, $\beta_k[m]$ is the large-scale channel coefficient that depends on the distance between the UAV and GT $k$ as given in \eqref{eq:betakm}, and $g_k[m,l]$ is a random variable with $\mathbb{E}[|g_k[m,l]|^2]=1$ accounting for the small-scale fading of the UAV-to-GT channel, which is independent and identically distributed (i.i.d.) for different $k$, $m$, $l$. Note that \eqref{eq:hkml} includes the LoS UAV-GT channel as a special case, for which $g_k[m,l]$ is deterministic with unit magnitude, i.e., $|g_k[m,l]|=1$.

\subsection{Problem Formulation}
 Denote by $P$ the transmission power of the UAV. The achievable rate in bps between the UAV and GT $k$ during the transmission of the $(m, l)$th packet is given by
 \begin{align}
 C_k[m,l]&=B \log_2\left(1+ \frac{P \big |h_k[m,l] \big|^2}{\sigma^2 \Gamma }\right)\notag \\
 &=B \log_2\left(1+ \frac{P\beta_k[m]\big|g_k[m,l]\big|^2}{\sigma^2 \Gamma}\right),
 \end{align}
 where $B$ denotes the channel bandwidth in Hertz (Hz), $\sigma^2$ represents the power of the additive white Gaussian noise (AWGN) at the GT receivers, and $\Gamma>1$ is the signal-to-noise ratio (SNR) gap between the practical modulation schemes and the theoretical Gaussian signaling. With the UAV's transmission rate fixed to $R$, the $(m,l)$th packet can be successfully received by GT $k$ if and only if $C_k[m,l]\geq R$. Thus, the probability that GT $k$ can successfully receive the $(m,l)$th packet can be expressed as
 \begin{align}
 p_k&[m,l]=\Pr \big(C_k[m,l] \geq R \big) \notag \\
 &=\Pr \left(\big|g_k[m,l]|^2 \geq \frac{\gamma_{\thr}}{\bar{\gamma}_0} \left( H^2 + \|\mathbf q[m] - \mathbf w_k\|^2\right)^{\alpha/2} \right) \notag \\
 &=F\left(\frac{\gamma_{\thr}}{\bar{\gamma}_0} \left( H^2 + \|\mathbf q[m] - \mathbf w_k\|^2\right)^{\alpha/2}  \right), \label{eq:pkml}
 \end{align}
 where $\gamma_{\thr}\triangleq 2^{R/B}-1$ is the SNR threshold for successful packet reception, $\bar{\gamma}_0\triangleq P\beta_0/(\sigma^2 \Gamma)$ is the average received SNR at the reference distance of 1m, and  $F(x)$ denotes the complementary cumulative distribution function (ccdf) of the random variable $|g_k[m,l]|^2$, which, by definition, is a non-increasing function with respect to $x$ for any given fading distribution. We assume that $F(x)$ is known in this paper. Define a distance parameter $D^*$ for the UAV-GT horizontal separation such that the average received SNR at $D^*$ equals $\gamma_{\thr}$, i.e., the resulting argument in \eqref{eq:pkml} equals $1$, we then have
 \begin{align}
 D^*=\sqrt{\left( \bar{\gamma}_0/{\bar {\gamma}_{\mathrm{th}}}\right)^{2/\alpha}-H^2}. \label{eq:Dstar}
 \end{align}
 For the special case of deterministic LoS channel such that $|g_k{[m,l]}|=1$, we have $F(x)=1$ if $x\leq 1$ and $0$ otherwise. In this case, we have $p_k[m,l]=1$ if $\|\mathbf q[m]-\mathbf w_k\|\leq D^*$ and $0$ otherwise. In other words, for the special case of LoS channel, a packet is guaranteed to be successfully received if the UAV-GT horizontal distance is no greater than $D^*$ and it will be lost otherwise. Note that for practical fading channels, all the $L$ packets transmitted by the UAV within the same time slot experience i.i.d. fading for any given GT $k$ and time slot $m$, since its distance from the UAV is assumed to be constant in each slot. Thus, $p_k[m,l]$ in \eqref{eq:pkml} is independent of $l$ but only depends on the slot number $m$.

 Let $Z_k[m,l], m=1,...,M, l=1,...,L$,  be a random variable indicating whether the $(m,l)$th packet is successfully received by GT $k$, which follows the Bernoulli distribution with success probability $p_k[m,l]$, denoted as $Z_k[m,l]\sim \mathrm {Bern}(p_k[m,l])$. The total number of packets that can be successfully received by GT $k$, denoted as $N_k$, is then a random variable given by
 \begin{align}
 N_k=\sum_{m=1}^{M}\sum_{l=1}^{L}Z_k[m,l], \ k\in \mathcal K.\label{eq:Nk}
  \end{align}
  Since $Z_k[m,l]$ are independent Bernoulli random variables with possibly different success probabilities, $N_k$ follows the Poisson binomial distribution \cite{929}. 

 Recall that with $N=ML$ network coded packets transmitted by the UAV, each GT is able to recover the information file as long as any $N'$ out of the $N$ packets are successfully received, whose probability can be written as
 \begin{align}
 P_{k, \succf}&\triangleq\Pr \big(N_k \geq N' \big), \ k\in \mathcal K. \label{eq:PkSucc}
 \end{align}

 Thus, the problem to minimize the mission completion time via trajectory optimization while ensuring a targeting file recovery probability $\bar P$ for all GTs can be formulated as
 \begin{align}
 \mathrm{(P1):}  \  \underset{\mathbf q[m], M}{\min} & \   M \notag \\
 \text{s.t.} & \  P_{k, \succf}\geq \bar P, \ \forall k\in \mathcal K, \label{eq:Constraint}\\
 & \ \big \|\mathbf q[m]-\mathbf q[m-1]\big\| \leq \tilde{V}_{\max}, \ m=2,\cdots, M.
 \end{align}

\section{Lower Bound of $P_{k, \succf}$ and Problem Reformulation}\label{sec:reFormul}
Problem $\mathrm{(P1)}$ is difficult to be directly solved. One major difficulty lies in that the successful file recovery probability $P_{k, \succf}$ in \eqref{eq:PkSucc} is related to the UAV trajectory $\{\mathbf q[m]\}$ in a rather implicit and complicated manner. In fact, even with a given UAV trajectory, and hence with known success probability $p_k[m,l]$ for each of the $N=ML$ transmitted packets, the complexity for evaluating the probability mass function (pmf) of $N_k$ is exponential with respect to $N$. This makes it quite challenging to obtain the optimal solution to $\mathrm{(P1)}$. In this paper, we propose an efficient approximate solution to $\mathrm{(P1)}$. To this end, we first derive an analytical lower bound for $P_{k, \succf}$ and transform the constraint \eqref{eq:Constraint} into a more tractable form in terms of the minimum  connection time between the UAV and each GT, during which their distance is below a certain threshold. We then propose effective trajectory designs for the reformulated optimization problem.

\begin{figure}
\centering
\includegraphics[scale=0.5]{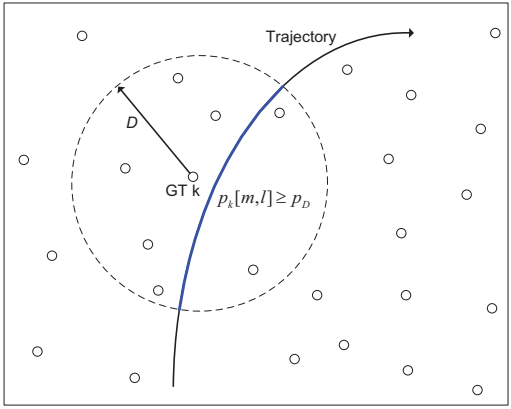}
\caption{Illustration of the lower bound derivation for $P_{k, \succf}$.}\label{F:LowerBoundIllustration}
\end{figure}

\subsection{Lower Bound of $P_{k, \succf}$}
As can be seen from \eqref{eq:pkml}, \eqref{eq:Nk} and \eqref{eq:PkSucc}, the successful file recovery probability $P_{k,\succf}$ for each GT $k$ is determined by the pmf of $N_k$, which in turn implicitly depends on the UAV trajectory $\mathbf q[m]$ via the successful packet reception probability $p_k[m,l]$. Due to UAV mobility, the packets transmitted by the UAV in different time slots in general experience non-identically distributed channels, i.e., $p_k[m,l]\neq p_k[m',l]$, $m\neq m'$. This makes it challenging to find an explicit expression for $P_{k,\succf}$ in terms of the UAV trajectory $\mathbf q[m]$ via directly deriving the pmf of $N_k$ in \eqref{eq:Nk}. To overcome this issue, we derive a lower bound for $P_{k, \succf}$ in \eqref{eq:PkSucc}, whose relationship with the UAV trajectory can be revealed more explicitly. As illustrated in Fig.~\ref{F:LowerBoundIllustration}, the main idea is to introduce an auxiliary distance parameter $D$, and ignore the portion of the UAV flight time during which the horizontal distance with each GT of interest is greater than $D$, hence incurring relatively higher packet loss probabilities than a threshold value (specified by $D$). Furthermore, for the considered time slots, the packet success probabilities are guaranteed to be no smaller than that corresponding to $D$, based on which a lower bound on the file recovery probability can be obtained. The detailed derivations are given as follows.

For any given auxiliary distance parameter $D\geq 0$, denote as $p_D$ the probability that a packet transmitted by the UAV is successfully received by a GT that has a horizontal distance $D$ from the UAV. Based on \eqref{eq:pkml}, for any channel model with known ccdf of the small-scale fading given by $F(\cdot)$, $p_D$ can be expressed as
\begin{align}\label{eq:pD}
p_D=F\left(\frac{\gamma_{\thr}}{\bar{\gamma}_0}\left(H^2+D^2\right)^{\alpha/2}\right).
\end{align}
Furthermore, for any UAV trajectory $\{\mathbf q[m]\}_{m=1}^M$, define the set $\mathcal M_{k,D}\subset \{1, \cdots, M\}$ for GT $k$ as the subset of all time slots such that the horizontal distance between the UAV and GT $k$ is no greater than $D$, i.e.,
\begin{align}
\mathcal M_{k,D} \triangleq \{m: \|\mathbf q[m]- \mathbf w_k\| \leq D\}. \label{eq:MkD}
\end{align}
For any given $D$, if $m\in \mathcal M_{k,D}$, we deem that the UAV and GT $k$ are {\it in connection} at time slot $m$; otherwise, they are not connected. Then the cardinality of $\mathcal M_{k,D}$, denoted as $|\mathcal M_{k,D}|$, is referred to as the number of connection time slots between UAV and GT $k$.
Since $F(\cdot)$ is a non-increasing function by definition,  based on \eqref{eq:pkml}, the following inequality holds for any given $D$,
\begin{align}
p_k[m,l]\geq p_D, \ \forall m\in \mathcal M_{k,D}. \label{eq:pkmlLB}
\end{align}
\begin{theorem}\label{theorem:PkLB}
For any given $D\geq 0$, the successful file recovery probability for GT $k$ defined in \eqref{eq:PkSucc} is lower-bounded as
\begin{align}
P_{k, \succf} \geq P_{k,\mathrm{lb}} \triangleq
\Pr \left( \hat {N}_k \geq N' \right), \label{eq:Pklb}
\end{align}
where $\hat N_k \sim \mathcal{B}\left(\big |\mathcal M_{k,D} \big |L, p_D\right)$ is a binomial random variable with $|\mathcal M_{k,D}|L$ independent trials each with success probability $p_D$. 
\end{theorem}
\begin{IEEEproof}
To prove Theorem~\ref{theorem:PkLB}, we need the following result.
\begin{lemma}\label{lemma:LBByBinomial}
Let $X_n\sim \mathrm{Bern}(p_n), n=1,\cdots, N$, be $N$ independent Bernoulli random variables with success probability $p_1,\cdots, p_N$, respectively. Then $X\triangleq \sum_{n=1}^N X_n$ follows a Poisson binomial distribution.  Furthermore, let $\hat X$ be a  binomial random variable with $\hat X\sim \mathcal{B}(N, \hat p)$ whose success probability satisfies $\hat p\leq p_n, \forall n$. Denote the ccdf of $X$ and $\hat X$ as $F_X(x)\triangleq \Pr(X\geq x)$ and $F_{\hat X}(x)\triangleq \Pr(\hat X\geq x)$, respectively. We then have
\begin{align}
F_X(x) \geq F_{\hat X}(x),  x=0, 1,\cdots, N. \label{eq:FXIneq}
\end{align}
\end{lemma}
\begin{IEEEproof}
Please refer to Appendix~\ref{A:LBByBinomial}.
\end{IEEEproof}

By substituting \eqref{eq:Nk} into \eqref{eq:PkSucc}, $P_{k, \succf}$ can be expressed as
\begin{align}
P_{k, \succf}& \triangleq \Pr \left(\sum_{m=1}^M \sum_{l=1}^L Z_k[m,l] \geq N' \right) \\
& \geq \Pr \left(\sum_{m\in \mathcal M_{k,D}} \sum_{l=1}^L Z_k[m,l]  \geq N' \right) \label{eq:ineq1} \\
& \geq \Pr \left( \hat N_k \geq N' \right)\triangleq  P_{k,\mathrm{lb}}, \label{eq:ineq2}
\end{align}
where \eqref{eq:ineq1} holds since $\mathcal M_{k,D}\subset \{1,\cdots, M\}$ for any $D\geq 0$, and \eqref{eq:ineq2} is obtained by applying Lemma~\ref{lemma:LBByBinomial} together with the inequality \eqref{eq:pkmlLB}.
\end{IEEEproof}

\subsection{Problem Reformulation}
With Theorem~\ref{theorem:PkLB}, for any chosen $D$, by replacing $P_{k, \succf}$ in \eqref{eq:Constraint} with its lower bound $P_{k,\mathrm{lb}}$, $\mathrm{(P1)}$ is recast into
 \begin{align}
 \mathrm{(P2):}  \  \underset{\mathbf q[m], M}{\min} & \  M \notag \\
 \text{s.t.} & \  P_{k,\mathrm{lb}} \geq \bar P, \ \forall k\in \mathcal K, \label{eq:ConstraintLB}\\
 & \ \big \|\mathbf q[m]-\mathbf q[m-1]\big\| \leq \tilde{V}_{\max}, \ m=2,\cdots, M.
 \end{align}
 Note that if \eqref{eq:ConstraintLB} is satisfied, then \eqref{eq:Constraint} is guaranteed to be satisfied as well due to the lower bound in \eqref{eq:Pklb}, but the reverse is not true in general. Therefore, for any given $D$, the optimal objective value of $\mathrm{(P2)}$ provides an upper bound to that of $\mathrm{(P1)}$. Thus, by solving $\mathrm{(P2)}$ for some appropriately chosen values for $D$, $\mathrm{(P1)}$ can be approximately solved. 
 As will be discussed in Section~\ref{sec:numerical}, one reasonable choice of $D$ is given by \eqref{eq:Dstar}.  In the following, we focus on solving $\mathrm{(P2)}$ for any given value of $D$.


 To obtain a more tractable form for the constraint \eqref{eq:ConstraintLB}, note that with moderately large $|\mathcal M_{k,D}|L$, the binomial random variable $\mathcal {B} (|\mathcal M_{k,D}|L, p_D)$ defined in Theorem~\ref{theorem:PkLB} can be well approximated by Gaussian random variable $\mathcal N(\mu, v^2)$, where $\mu=|\mathcal M_{k,D}|Lp_D$ and $v^2=|\mathcal M_{k,D}|Lp_D(1-p_D)$. As a result, the lower bound $P_{k,\mathrm{lb}}$ defined in \eqref{eq:Pklb} can be approximated as
 \begin{align}
 P_{k,\mathrm{lb}}\approx Q\left(\frac{N'-|\mathcal M_{k,D}|Lp_D}{\sqrt{|\mathcal M_{k,D}|Lp_D(1-p_D)}}\right), \label{eq:QApprox}
 \end{align}
 where $Q(x)\triangleq \int_0^\infty e^{-u^2/2}du$ is the Gaussian Q-function. Therefore, by substituting \eqref{eq:QApprox} into constraint \eqref{eq:ConstraintLB} and solving for $|\mathcal M_{k,D}|$, we get
 \begin{align}
 |\mathcal M_{k,D}| \geq M_{\min} \triangleq A^2/L, \label{eq:MkCardConstr}
 \end{align}
 where
 \begin{equation}
 \begin{aligned}
 \vspace{-5ex} A\triangleq \frac{1}{2 \sqrt{p_D}}\left(\sqrt{4N'+(1-p_D)(Q^{-1}(\bar P))^2}-Q^{-1}(\bar P)\sqrt{1-p_D}\right), \label{eq:J}
 \end{aligned}
 \end{equation}
 with $Q^{-1}(\cdot)$ denoting the inverse Gaussian Q-function.

In other words, for any given $D$, the constraint \eqref{eq:ConstraintLB} on the success file recovery probability is equivalent to the constraint that the number of connection time slots $|\mathcal M_{k,D}|$ between the UAV and each GT should be no smaller than the minimum threshold $M_{\min}$, where $M_{\min}$ is a constant determined by $p_D$, $\bar P$ and $N'$. To gain more insights for \eqref{eq:MkCardConstr}, consider the special case when $D$ is sufficiently small such that $p_D\rightarrow 1$. In this case, it follows from \eqref{eq:MkCardConstr} that we have $M_{\min}=N'/L$. In other words, if $D$ is small so that each packet transmitted by the UAV can be successfully received almost surely by those GTs in connection with the UAV, then the UAV only needs to stay in connection with each GT for $N'/L$ time slots to transmit $N'$ packets, as expected. On the other hand, if $D$ is chosen to be large such that $p_D\rightarrow 0$, it then follows from \eqref{eq:MkCardConstr} and \eqref{eq:J} that we have $M_{\min} \propto 1/p_D$, i.e., the minimum number of connection time slots $M_{\min}$ increases inversely proportional with $p_D$.

Define the following indicator function
\begin{align}
I_{k,D}[m]=\begin{cases}
1, \ & \text{if }  \|\mathbf q[m]-\mathbf w_k \| \leq D, \\
0, \ & \text{otherwise}.
\end{cases}
\end{align}
Then $|\mathcal M_{k,D}|=\sum_{m=1}^M I_{k,D}[m]$. Therefore, $\mathrm{(P2)}$ can be reformulated as
 \begin{align}
   \underset{\mathbf q[m], M}{\min} & \   T=\delta_t M \notag \\
 \text{s.t.} & \  |\mathcal M_{k,D}|\geq M_{\min}, \ \forall k\in \mathcal K, \label{eq:ConstraintLB2}\\
 & \ \big \|\mathbf q[m]-\mathbf q[m-1]\big\| \leq \tilde{V}_{\max}, \ m=2,\cdots, M.
 \end{align}
 When the time slot length $\delta_t$ is chosen to be sufficiently small, then the above problem can be written in its continuous-time format as
  \begin{align}
 \mathrm{(P3):}  \  \underset{\mathbf q(t), T}{\min} & \   T \notag \\
 \text{s.t.} & \   T_{k,D}\triangleq \int_0^T I_{k,D}(t)dt \geq T_{\min}, \forall k\in \mathcal K \label{eq:ConstraintLB3}\\
& \ \|\dot{\mathbf q}(t)\| \leq V_{\max},  0\leq t \leq T, \label{eq:speedConstr}
 \end{align}
 where $T_{\min}\triangleq M_{\min} \delta_t$ and
 \begin{align}\label{eq:Ikt}
 I_{k,D}(t)=\begin{cases}
1, \ & \text{if }  \|\mathbf q(t)-\mathbf w_k \| \leq D, \\
0, \ & \text{otherwise}.
\end{cases}
 \end{align}
 In the next section, we focus on solving the trajectory optimization problem $\mathrm{(P3)}$. 

\section{Proposed Trajectory Design}\label{sec:PropSol}
The main challenge for optimally solving $\mathrm{(P3)}$ lies in the non-convex constraint \eqref{eq:ConstraintLB3}, which involves time-dependent indicator functions \eqref{eq:Ikt} in terms of the UAV trajectory. To solve $\mathrm{(P3)}$, we first show the following result.

\begin{theorem}\label{theorem:LineSegments}
Without loss of optimality to $\mathrm{(P3)}$, the UAV trajectory can be assumed to constitute only connected line segments.
\end{theorem}
\begin{IEEEproof}
Please refer to Appendix~\ref{A:LineSegments}.
\end{IEEEproof}

Theorem~\ref{theorem:LineSegments} implies that finding the optimal solution to $\mathrm{(P3)}$ is equivalent to finding the optimal set of ordered waypoints $\mathcal Q_{\mathrm{wp}}$, which contains the locations representing the starting and ending points of each line segment, as well as optimizing the instantaneous UAV speed along the path connecting the waypoints. 
 However, finding the optimal set of waypoints $\mathcal Q_{\mathrm{wp}}$ is a challenging problem in general.  In fact, for the extreme case when $D=0$, the constraint \eqref{eq:ConstraintLB3} reduces to that the UAV needs to sequentially visit all the $K$ GTs and stay stationary on top of each for at least $T_{\min}$ seconds. In this case, finding the optimal waypoints to $\mathrm{(P3)}$ reduces to determining the visiting order of all the $K$ GTs so as to minimize the total UAV travelling distance, which is essentially equivalent to the classic TSP \cite{TSP, 909, 908}. The only difference is that different from the standard TSP, the traveller/UAV in our considered problem does not need to return to the origin where it starts the tour. Note that TSP is an NP-hard problem in combinatorial optimization. However, various heuristic and high-quality approximation algorithms have been developed. A brief overview on TSP and its variations are given in Appendix~\ref{A:TSP}. On the other hand, for the general case with $D>0$, $\mathrm{(P3)}$ seems to be similar to the TSP with neighborhoods (TSPN) \cite{927}. However, as existing algorithms for TSPN such as \cite{928} assume that the neighborhoods are disjoint disks and do not have the minimum connection time constraints, they cannot be directly applied for solving problem $\mathrm{(P3)}$.  In the following, for $\mathrm{(P3)}$ with the general $D\geq 0$, we will first present a simple benchmark scheme by taking the GTs as the waypoints, and then propose two more efficient schemes for waypoints design based on a novel concept of VBS placement and by applying convex optimization techniques. Furthermore, for any given waypoints design, the optimal UAV speed over time will be efficiently obtained via solving an LP problem.




\subsection{Waypoint Design} \label{sec:wayPointDesign}
{\it (1) Scheme 1 (benchmark): GTs as Waypoints.} Note that a feasible UAV trajectory to $\mathrm{(P3)}$ needs to ensure that the minimum connection time constraints in \eqref{eq:ConstraintLB3} are satisfied with the designed waypoints. For any $D\geq 0$, one straightforward approach to ensure the feasibility of \eqref{eq:ConstraintLB3} is to let the UAV sequentially visit (i.e., stay on top of) all GTs. More specifically, $\mathcal Q_{\mathrm{wp}}$ is determined by simply applying the TSP algorithm over all the $K$ GTs (without the need of returning to the origin as discussed in Appendix~\ref{A:TSP}). 
In this case, since each GT is guaranteed to be in connection with the UAV when it is just above the GT, the constraints in \eqref{eq:ConstraintLB3} can be met by appropriate UAV speed optimization, as will be studied in Section~\ref{sec:SpeedOpt}.

{\it (2) Scheme 2 (proposed): VBSs as Waypoints.} It is intuitive to see that for a given $D>0$, it is in general unnecessary for the UAV to fly over all the GTs since at one location, the UAV could be in connection with more than one GTs simultaneously. Thus, the number of waypoints that the UAV needs to visit to ensure the feasibility of \eqref{eq:ConstraintLB3} could be much less than $K$, especially when $D$ is large and the GTs are densely distributed. Therefore, in this subsection, we propose an alternative waypoint design based on a new idea of VBS placement.

Specifically, given the GT locations $\{\mathbf w_k\}$ and the UAV threshold coverage range $D$, the VBS placement problem aims to find a minimum number of VBSs and their respective locations, so that each GT is covered by at least one VBS. This problem resembles the standard BS placement problem for ensuring user coverage with a given coverage distance $D$, where several efficient algorithms have been proposed, such as the spiral BS placement algorithm proposed in \cite{886}. Let $G \leq K$ be the minimum number of VBSs obtained by applying the BS placement algorithm, and their locations are denoted as $\mathbf v_g \in \mathbb{R}^{2\times 1}$, $g=1,\cdots, G$. An efficient waypoint design to ensure the feasibility of \eqref{eq:ConstraintLB3} is to let the UAV sequentially visit these VBSs by following the path obtained by the TSP algorithm applied over $\{\mathbf v_g\}_{g=1}^G$. In this case, the number of waypoints that the UAV needs to travel is $G$, which is in general less than $K$.

\begin{figure}
\centering
\includegraphics[scale=0.5]{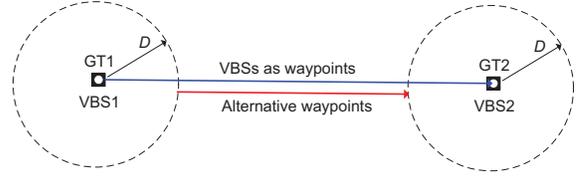}
\caption{A toy example for illustrating the inefficacy of directly using VBSs as waypoints.}\label{F:Example}
\end{figure}

{\it (3) Scheme 3 (proposed): Waypoints Based on VBS Placement and Convex Optimization.} Traversing over all the $G$ VBSs, though providing a feasible waypoints design to $\mathrm{(P3)}$, may not always be desirable. This is illustrated by a toy example shown in Fig.~\ref{F:Example}, where there are two GTs, each covered by one VBS that is placed in essentially the same location as the GT. It is observed that traversing over both VBSs in fact leads to unnecessarily longer trajectory than the alternative design shown in Fig.~\ref{F:Example}. To overcome this limitation, in this subsection, we propose a more efficient waypoint design based on the placed VBSs and by applying convex optimization techniques.

Specifically, with VBS placement and TSP algorithm applied over the obtained $G$ VBSs, the GTs in $\mathcal K$ are essentially partitioned into $G$ {\it ordered} clusters $\mathcal S_g$, $g=1,\cdots, G$, where $\mathcal S_g\subset \mathcal K$ denotes the subset of GTs that are covered by the $g$th VBS while applying the VBS placement algorithm. For the $g$th ordered cluster with GTs $\mathcal S_g$, define the following set
\begin{align}
\mathcal C_g \triangleq \{\mathbf q\in \mathbb{R}^{2\times 1}:  \| \mathbf q - \mathbf w_k \| \leq D, \forall k \in \mathcal S_g \}.\label{eq:Cm}
\end{align}
In other words, $\mathcal C_g$ is the set of all possible UAV locations ensuring that all GTs in $\mathcal S_g$ are simultaneously in connection with the UAV. It is obvious that $\mathcal C_g$ is non-empty (since the VBS $g$ with location $\mathbf v_g$ belongs to this set) and a convex set (since it is an intersection of $|\mathcal S_g|$ convex sets). As a result, as long as the UAV sequentially visits all the $G$ convex regions $\mathcal C_g$, the constraints in \eqref{eq:ConstraintLB3} can be met by appropriate UAV speed optimization. In the following, the waypoints in each of the convex region $\mathcal C_g$ is optimized.

Without loss of generality, let $\mathbf s_g, \mathbf f_g \in \mathcal C_g$ be the starting and ending points of the UAV trajectory intersecting with the region $\mathcal C_g$, respectively. Note that since $\mathcal C_g$ is a convex set, all points on the line segment between $\mathbf s_g$ and $\mathbf f_g$ are also in $\mathcal C_g$, i.e., they ensure that all the GTs in $\mathcal S_g$ are in connection with the UAV. Given the UAV's maximum flying speed $V_{\max}$, the minimum time required for the UAV to travel within the region $\mathcal C_g$, i.e., from $\mathbf s_g$ to $\mathbf f_g$, is $\frac{\|\mathbf f_g - \mathbf s_g\|}{V_{\max}}$. On the other hand, to ensure the minimum connection time constraint \eqref{eq:ConstraintLB3}, one viable approach is to ensure that the UAV remains in $\mathcal C_g$ for at least $T_{\min}$ seconds. Thus, the minimum time required for the UAV to travel within $\mathcal C_g$ is $\max \left\{\frac{\|\mathbf f_g - \mathbf s_g\|}{V_{\max}}, T_{\min}\right\}$. Furthermore, the minimum time required for the UAV to travel between $\mathcal C_g$ and $\mathcal C_{g+1}$ is $\frac{\|\mathbf s_{g+1}-\mathbf f_g\|}{V_{\max}}$. As a result, the waypoints $\{\mathbf s_g, \mathbf f_g\}_{g=1}^G$ could be designed by solving the following problem
\begin{align}
\mathrm{(P4):}  \  \underset{\{\mathbf s_g, \mathbf f_g\}_{g=1}^G}{\min} & \   \sum_{g=1}^G \max \left\{\frac{\|\mathbf f_g -\mathbf s_g\|}{V_{\max}}, T_{\min} \right\} + \notag \\
& \hspace{5ex} \sum_{g=1}^{G-1} \frac{\|\mathbf s_{g+1}-\mathbf f_g\|}{V_{\max}} \notag \\
 \text{s.t.} & \  \mathbf s_g, \mathbf f_g \in \mathcal C_g, \ \forall g.
\end{align}
Note that the cost function of $\mathrm{(P4)}$ is the total mission completion time with waypoints $\{\mathbf s_g, \mathbf f_g\}$, which is a convex function with respect to $\{\mathbf s_g, \mathbf f_g\}$. Furthermore, all the constraints in $\mathrm{(P4)}$ are convex. Thus, $\mathrm{(P4)}$ is a convex optimization problem, which can be efficiently solved by standard convex optimization techniques or  existing software such as CVX \cite{227}.

Note that as compared to the previous scheme by directly taking the VBSs as waypoints, the new waypoints obtained in $\mathrm{(P4)}$ avoid the unnecessary traveling to the VBSs, and thus are expected to achieve better performance, as will be numerically verified in Section~\ref{sec:numerical}.

\subsection{UAV Speed Optimization}\label{sec:SpeedOpt}
For any given set of feasible waypoints $\mathcal Q_{\mathrm{wp}}$, the UAV path is determined by sequentially connecting the waypoints $\mathcal Q_{\mathrm{wp}}$ with line segments. As a result, problem $\mathrm{(P3)}$ reduces to finding the optimal instantaneous UAV speed over time along the path connecting these waypoints. 
 To this end, we discretize the UAV path with the infinitesimal displacement $\delta_d$ (instead of over time) to get $J$ UAV sampled locations on the path, denoted by $\{\mathbf q_j\}_{j=1}^J$. 
 As a result, the corresponding value of the indicator function in \eqref{eq:Ikt} can be obtained, which is denoted as $I_{kj}$, $k\in \mathcal K$, $j=1,\cdots, J$. That is, $I_{kj}=1$ represents that the UAV is in connection with GT $k$ when it is at location $j$. Denote by $\tau_j \geq 0$ the time for the UAV to travel from location $\mathbf q_j$ to $\mathbf q_{j+1}$, with the speed $V_j=\frac{\delta_d}{\tau_j}$. Note that since $\delta_d$ is set sufficiently small, $V_j$ well approximates the instantaneous UAV speed, and we must have $\frac{\delta_d}{\tau_j}\leq V_{\max}$. 
For any given set of feasible waypoints, $\mathrm{(P3)}$ reduces to optimizing the UAV speed $V_j$ or equivalently the time duration $\tau_j$, $j=1,\cdots, J$, which is formulated as
\begin{align}
\mathrm{(P5):}  \  \underset{\{\tau_j\}}{\min} & \   \sum_{j=1}^J \tau_j \notag \\
 \text{s.t.} & \  \sum_{j=1}^J I_{kj}\tau_j \geq T_{\min}, \ \forall k\in \mathcal K, \\
 & \ \tau_j \geq \frac{\delta_d}{V_{\max}}, \ j=1,\cdots, J.
\end{align}
Note that $\mathrm{(P5)}$ is feasible if and only if $\forall k\in \mathcal K$, there exists at least one UAV location $j$ such that $I_{kj}=1$. This is guaranteed based on the three waypoint designs presented in Section~\ref{sec:wayPointDesign}. $\mathrm{(P5)}$  is a standard LP problem, which can be efficiently solved via e.g. \cite{227}.

\begin{table}
\centering
\caption{System setup for numerical simulations.}\label{table:setupNumerical}
{\renewcommand{\arraystretch}{1.2}\begin{tabular}{p{4cm}|p{4cm}}
\hline
UAV altitude & $H=100$m\\
\hline
Maximum UAV speed & ${V}_{\max}=50$m/s\\
\hline
UAV transmission power & $P=10$dBm \\
\hline
Bandwidth & $B=1$ MHz \\
\hline
Noise power  & $\sigma^2=-109$dBm \\
\hline
SNR gap & $\Gamma=10$ dB\\
\hline
Information file size & $W=2$ Mbits \\
\hline
Packet size & $R_p=10^4$ bits/packet \\
\hline
Minimum  number of packets required for file recovery & $N'=200$ \\
\hline
UAV transmission rate & $R=1$ Mbits/second \\
\hline
Time for transmitting one packet & $T_p=0.01$ seconds \\
\hline
Time slot length & $\delta_t=0.1$ seconds \\
\hline
Number of transmitted packets per slot & $L=10$\\
\hline
Channel gain at reference distance & $\beta_0=-40$ dB\\
\hline
Path loss exponent & $\alpha=2.6$ \\
\hline
Rician factor & $K_c=2$ \\
\hline
Target probability for file recovery & ${\bar P}=0.9$\\
\hline
\end{tabular}}
\end{table}

\renewcommand{\arraystretch}{1}

\begin{figure*}
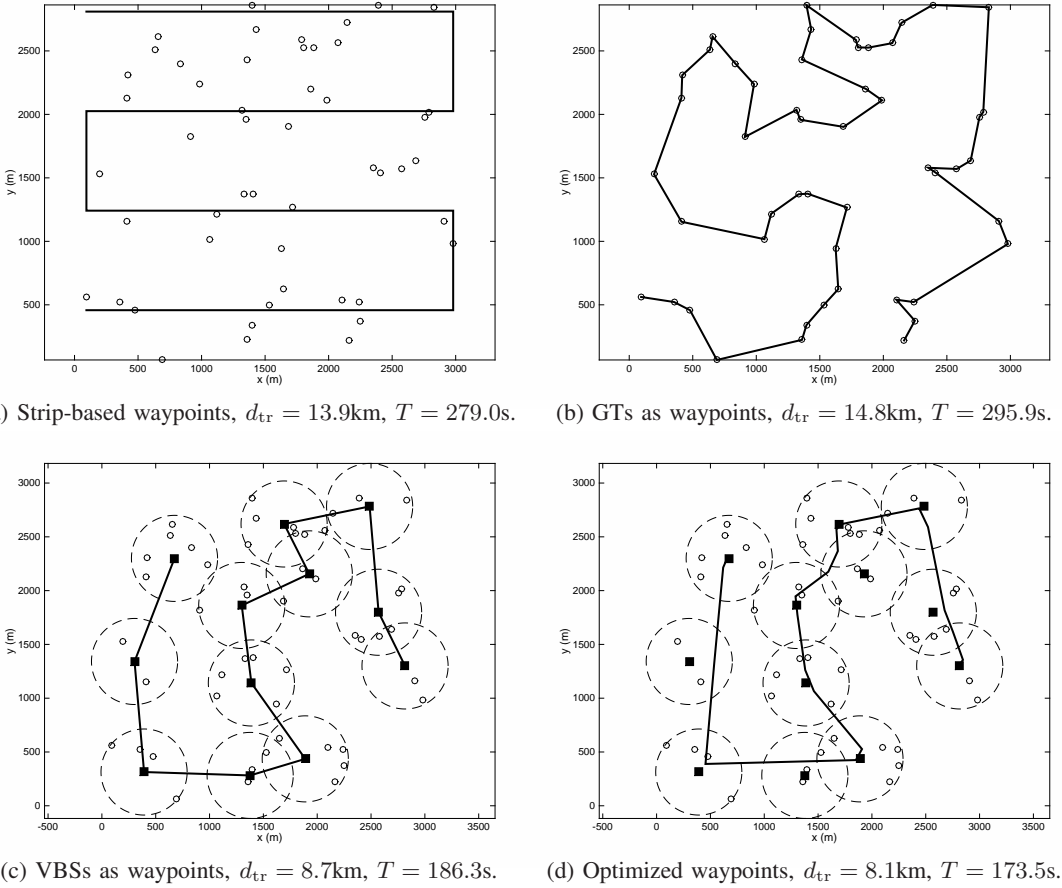

\centering
\begin{subfigure}{0.4\textwidth}
\centering
\includegraphics[width=0.9\linewidth]{TrajectoryFullAreaCoverage.eps}
\caption{Strip-based waypoints, $d_{\mathrm{tr}}=13.9$km, $T=279.0$s.}\label{F:TrajectoryFullAreaCoverage}
\end{subfigure}
\begin{subfigure}{0.4\textwidth}
\centering
\includegraphics[width=0.9\linewidth]{TrajectoryGTAsWayPoints.eps}
\caption{GTs as waypoints, $d_{\mathrm{tr}}=14.8$km, $T=295.9$s.}\label{F:TrajectoryGTAsWayPoints}
\end{subfigure}
\\
\vspace{3ex}
\begin{subfigure}{.4\textwidth}
\centering
\includegraphics[width=0.9\linewidth]{TrajectoryVBSAsWayPoints.eps}
\caption{VBSs as waypoints, $d_{\mathrm{tr}}=8.7$km, $T=186.3$s.}\label{F:TrajectoryVBSAsWayPoints}
\end{subfigure}
\begin{subfigure}{.4\textwidth}
\centering
\includegraphics[width=0.9\linewidth]{ProposedTrajectoryOptimizedWayPointsD400m.eps}
\caption{Optimized waypoints, $d_{\mathrm{tr}}=8.1$km, $T=173.5$s.}\label{F:ProposedTrajectoryOptimizedWayPointsD400m}
\end{subfigure}
\caption{Comparison of the UAV trajectories with different waypoint designs. Small circles denote GTs and squares represent VBSs.}\label{F:TrajectoryComparison}
\end{figure*}

\section{Numerical Results}\label{sec:numerical}
In this section, numerical results are provided to evaluate the performance of our proposed trajectory designs. We assume that the $K$ GTs are randomly and uniformly distributed in a square area of side length equal to $3000$m. For UAV-to-ground channels, 
 we adopt the practical Rician fading channel model, which is characterized by the Rician factor $K_c$ representing the power ratio between the LoS signal component to the scattered component. In this case, the small-scale fading coefficients $g_{k}[m,l]$ in \eqref{eq:hkml} can be explicitly modeled as
\begin{align}
g_k[m,l]&=\sqrt{\frac{K_c}{K_c+1}} \bar g + \sqrt{\frac{1}{K_c+1}}\tilde g\\
&=\sqrt{\frac{1}{2(K_c+1)}}\underbrace{\left(\sqrt{2K_c} \bar g + \sqrt{2} \tilde g \right)}_{Y}, \label{eq:gkmlRician}
\end{align}
where $\bar g$ denotes the deterministic LoS channel component with $|\bar g|=1$, and $\tilde g$ represents the random scattered component, which is a zero-mean unit-variance circularly symmetric complex Gaussian (CSCG) random variable. With $Y$ defined in \eqref{eq:gkmlRician}, $|Y|^2$ follows the non-central chi-square distribution with two degrees of freedom (DoF) and non-centrality parameter $\lambda=2K_c$, denoted as $|Y|^2 \sim \chi'^2_2(2K_c)$. Thus, the ccdf of $|g_k[m,l]|^2$ in \eqref{eq:pkml} can be explicitly written as
\begin{align}
F(z) & \triangleq \Pr(|g_k[m,l]|^2\geq z)=\Pr\big(|Y|^2\geq 2(K_c+1)z\big) \notag \\
& =Q_1\left(\sqrt{2K_c}, \sqrt{2(K_c+1)z} \right),\label{eq:FzRician}
\end{align}
where $Q_1(a,b)$ is the standard Marcum-Q-function. Unless otherwise stated, the numerical setup of the following simulations is given in Table~\ref{table:setupNumerical}.

For the proposed waypoint designs with VBSs placement, we use the spiral BS placement algorithm proposed in \cite{886} to obtain the VBSs. Furthermore, since the TSP problem involved in our design does not require the UAV to return to the starting point, we apply the strategy by adding a dummy node as described in Appendix~\ref{A:TSP}. The resulting TSP is solved by using the existing Matlab codes available in \cite{908}. Note that by applying the corresponding TSP variations as discussed in Appendix~\ref{A:TSP}, our proposed UAV trajectory design can be directly applied to the case when the UAV's initial and/or final locations are predetermined. Such extensions are omitted for brevity. Besides the three waypoint design schemes presented in Section~\ref{sec:wayPointDesign}, we also consider another benchmark scheme, called ``strip-based waypoints'', where the UAV's trajectory is designed to ensure full area coverage. Specifically, for any given realization of the GT locations and chosen distance parameter $D$, the UAV first obtains the smallest rectangle that contains all the $K$ GTs, and then partitions this rectangular area into rectangular strips each with width $2D$. The UAV then sequentially travels along the center of the rectangular strips, as shown in Fig.~\ref{F:TrajectoryComparison}(a). Note that such a trajectory design ensures that all locations within the rectangular area are covered by the UAV. For all the four trajectory design schemes, the UAV's instantaneous speed is optimized based on the LP problem $\mathrm{(P5)}$, given their respective waypoints.

\subsection{Trajectory Comparison and Lower Bound Verification}
By choosing the auxiliary distance parameter as $D=400$m, Fig.~\ref{F:TrajectoryComparison} compares the different UAV trajectories with the four considered waypoint designs for one specific realization of the GT locations with $K=50$. The corresponding total UAV traveling distances $d_{\mathrm{tr}}$ and the mission completion time $T$ are also shown in the figure. It is observed that for both benchmark schemes with strip-based waypoints and GTs as waypoints,  the UAV needs to travel  longer distances and hence require larger mission completion time, as compared to the proposed designs as shown in Fig.~\ref{F:TrajectoryComparison}(c) and Fig.~\ref{F:TrajectoryComparison}(d).
 This is expected since compared to the two benchmark schemes, the proposed designs jointly utilize the information of the GT locations and the coverage distance $D$ via VBS placement and convex optimization. Furthermore, by comparing Fig.~\ref{F:TrajectoryComparison}(c) and Fig.~\ref{F:TrajectoryComparison}(d), it is observed that by solving the convex optimization problem $\mathrm{(P4)}$ based on the obtained VBSs, the UAV can further reduce its required traveling distance and mission completion time by avoiding the unnecessary visit to all the VBSs.

\begin{figure}
\centering
\includegraphics[scale=0.4]{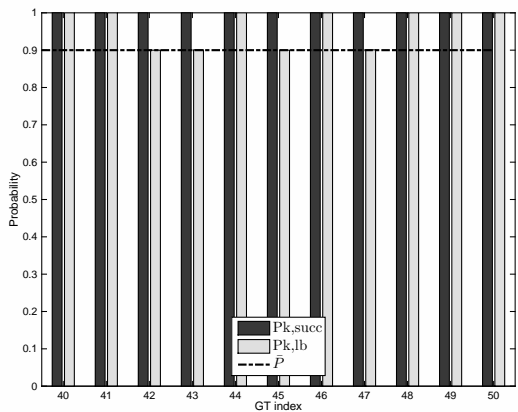}
\caption{Numerical verification of the lower bound for the succuss file recovery probability.}\label{F:LBVerification}
\end{figure}

For the proposed UAV trajectory shown in Fig.~\ref{F:TrajectoryComparison}(d), Fig.~\ref{F:LBVerification} plots the actual file recovery probability $P_{k,\succf}$ and our derived lower bound $P_{k,\lb}$, where $P_{k,\succf}$ is obtained numerically via Monte Carlo simulations over $10^4$ random channel realizations. Note that for better visualization, only the results for 10 of the GTs are shown in the figure. It is observed that with the proposed UAV trajectory design, the constraints in \eqref{eq:ConstraintLB} based on the lower bound of the file recovery probability are satisfied with strict equality for some of the GTs, as expected. Furthermore, it is found that with the optimized UAV trajectory, all GTs are able to successfully recover the file almost surely, i.e., with actual success probability almost equal to 1. This verifies the proposed lower bound and also shows the effectiveness of the proposed trajectory design.


\subsection{Effect of Auxiliary Distance Parameter $D$}
\begin{figure}
\centering
\includegraphics[scale=0.45]{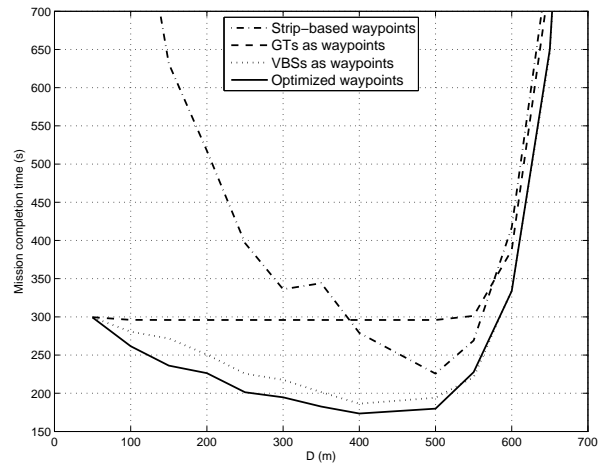}
\caption{Mission completion time versus $D$.}\label{F:MissionComTimeVsD}
\end{figure}

Next, we study the effect of the auxiliary distance parameter $D$ on the system performance.
 Fig.~\ref{F:MissionComTimeVsD} plots the total mission completion time versus $D$ for the four UAV trajectory design schemes, with the  GT locations same as Fig.~\ref{F:TrajectoryComparison}. It is observed that for all schemes, the mission completion time has the general trend of firstly decreasing and then increasing with $D$. This is expected since the value of $D$ affects the UAV trajectory design in two different ways. On one hand, increasing $D$ leads to lower successful packet reception probability $p_D$ in \eqref{eq:pD}, which in turn requires that each GT to keep in connection with the UAV for a longer duration in order to ensure the same file recovery probability. From this perspective, the mission completion time tends to increase with $D$. On the other hand, as $D$ increases, there will be more GTs that are simultaneously in connection with the UAV. As a result, the UAV in general needs to travel shorter distances if larger $D$ is chosen. From this perspective, the mission completion time tends to decrease with the increasing of $D$. Thus, for any given GT locations, there exists an optimal threshold distance $D$ that balances the above two conflicting effects and achieves the minimum mission completion time. To our best effort, it is challenging to find the optimal value $D$ analytically. However, as illustrated in Fig.~\ref{F:MissionComTimeVsD}, one good choice for $D$ is  such that the average received SNR when the GT and UAV are separated by horizontal distance $D$ is equal to the threshold SNR $\bar {\gamma}_{\mathrm{th}}$, in which case $D$ is given by $D^*$ in \eqref{eq:Dstar}. For the setup under consideration, we have $D^*=430.3$m, which gives the near optimal choice based on Fig.~\ref{F:MissionComTimeVsD}.


\subsection{Performance Comparison}
\begin{figure}
\centering
\includegraphics[scale=0.45]{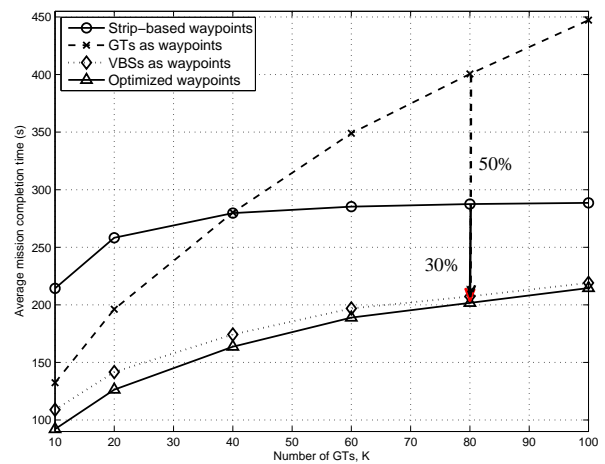}
\caption{Average mission completion time versus the number of GTs.}\label{F:AvgMissionCompletionTimeVsK}
\end{figure}

Fig.~\ref{F:AvgMissionCompletionTimeVsK} compares the average mission completion time versus the number of GTs $K$, where the average is taken over $100$ random realizations of the GT locations. For all schemes, the auxiliary distance parameter $D$ is set as $D^*=430.3$m. It is first observed that for small or moderate number of GTs, all the three trajectories with the waypoints designs given in Section~\ref{sec:wayPointDesign} significantly outperform the benchmark strip-based trajectory. This is expected since when the GTs are sparsely distributed, utilizing the location information of the GTs more wisely is beneficial for the UAV trajectory design. As $K$ increases or the GTs are more densely deployed, the trajectory design by simply taking the GTs as the waypoints performs worse than the other benchmark scheme with strip-based waypoints, since it becomes time wasteful for the UAV to visit all the GTs even when many of them are near to each other. For all the $K$ values considered, both proposed designs with the VBSs as waypoints or with the optimized waypoints significantly outperform the two benchmark schemes. For instance, for $K=80$, the mission completion time with the two proposed trajectory designs is reduced by around $50\%$ as compared to the benchmark scheme with GTs as waypoints, and by $30\%$ than the strip-based waypoints design. 

%

\begin{figure}
\centering
\includegraphics[scale=0.45]{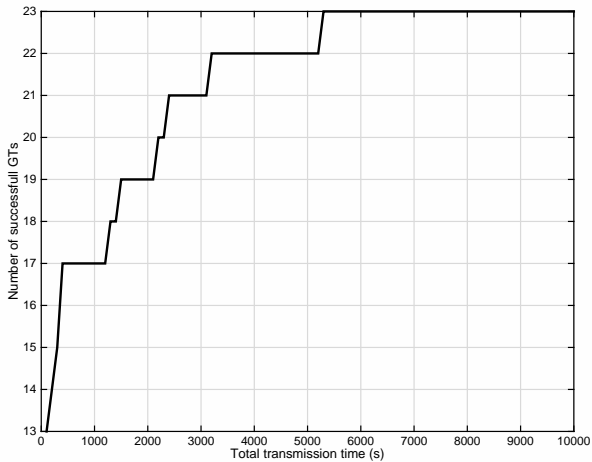}
\caption{Number of successful GTs versus transmission time for the benchmark scheme with a static transmitter.}\label{F:NumberSuccNodesStatic}
\end{figure}

Last, to illustrate the performance gain by exploiting the high UAV mobility, we consider another benchmark multicasting scheme with static transmitter, i.e., the horizontal projection of the transmitter (e.g., a static UAV) is fixed at the geometric center of the GTs. For $K=100$ GTs, 
Fig.~\ref{F:NumberSuccNodesStatic} shows the number of successful GTs (i.e., those with successful file recovery probability no smaller than $\bar P$) for the benchmark scheme with a static transmitter as the transmission time increases. 
 It is observed that although the number of successful GTs increases with the transmission time, or equivalently with the number of transmitted coded packets, the increasing rate is very slow.
 For example, even with the transmission time increased to $10^4$ s, only $23$ GTs are able to achieve the targeting file recovery probability. This is expected since with the transmitter fixed in location, the GTs that have a long distance with the transmitter suffer from high packet loss probabilities. On the other hand, with UAV-enabled multicasting with the proposed trajectory design, only about $210$s is needed to ensure that all the $100$ GTs satisfy the file recovery requirement, as can be seen from Fig.~\ref{F:AvgMissionCompletionTimeVsK}. This demonstrates the dramatic performance gain by exploiting the high mobility of UAVs for wireless multicasting.

\section{Conclusion}\label{sec:Conl}
This paper studied the trajectory design problem for a UAV-enabled multicasting system to minimize the mission completion time, while ensuring that each GT is able to successfully recover the file with a high target probability. We first converted the formulated optimization problem into a more tractable form based on the derived analytical lower bound of the successful file recovery probability, so that its complicated constraint for each GT is simplified to one on its  minimum connection time with the UAV. We showed that the optimal UAV trajectory only needs to constitute connected line segments, which  can be determined by finding the optimal set of waypoints and then the optimal speed over time along the path connecting the waypoints. We proposed two practical waypoints design schemes and applied the LP to find the optimal traveling speed given waypoints. Numerical results demonstrated significant performance gains of the proposed designs over various benchmark schemes.

\appendices

\section{Overview of Travelling Salesman Problem and Its Variations}\label{A:TSP}
In this section, we give a brief description on the classic TSP  \cite{TSP, 908, 909} and discuss its variations. The standard TSP is described as follows. Given a set of $K$ cities and the distances between each pair of the cities, a traveler wishes to start and end at the same city and visit each other city exactly once. The problem is to find the route (or sequence of visiting) such that the total traveling distance is minimized. TSP is an NP-hard problem in combinatorial optimization and hence is difficult to be optimally solved. 
 Various heuristic and approximation algorithms have been proposed to give efficient high-quality solutions \cite{TSP}, \cite{907}. In particular, the TSP  can be formulated as a binary integer programming, and an efficient high-quality solution can be obtained by using the existing Matlab optimization toolbox (Matlab version 2014 onwards). The complete Matlab codes and one illustrative example can be found in \cite{908}. On the other hand, for many applications, different variations of the TSP  need to be considered. In the following, we discuss five of these variations depending on whether the traveler needs to return to the origin and whether the origin/end city is predetermined.
\begin{enumerate}
\item {\it Return-GivenOrigin:} In this setup, the traveler needs to return to the origin city, and the origin/end city is predetermined. This is essentially the same as the standard TSP, which would return a closed tour so that any city can be regarded as the origin city.
\item {\it NoReturn-ArbitraryOriginAndEnd:} In this setup, the traveler does not need to return to the origin city, and the origin and end cities are not predetermined and hence can be optimized. The optimal solution can be found as follows \cite{909}. First, add a dummy city whose distances to all the existing $K$ cities are 0 (this is a virtual node that does not exist physically). Then solve the standard TSP problem for the $K+1$ cities, and then remove the two edges associated with the dummy city. It can be shown by contradiction that such a solution is optimal.
\item {\it NoReturn-GivenOriginAndEnd:} In this setup, the traveler does not need to return to the origin city, and the origin and end cities are both predetermined, denoted as $A$ and $B$, respectively. To solve this problem, we similarly add a dummy city, with its distance to both $A$ and $B$ set to $0$, whereas that to all other $K-2$ cities set to a sufficiently large number (so as to avoid the traveling from the dummy city to all other cities except $A$ and $B$). By solving the standard TSP problem for the $K+1$ cities, and then removing the two edges associated with the dummy city, we obtain the optimal solution.
\item {\it NoReturn-GivenOrigin-ArbitraryEnd:}  In this setup, the traveler does not need to return to the origin city, and only the origin city is predetermined, denoted as $A$. To solve this problem, we similarly add a dummy city whose distance to $A$ is set to $0$, whereas that to all other $K-1$ cities are set to an identical arbitrary positive value. By solving the standard TSP problem for the $K+1$ cities, and then removing the two edges associated with the dummy city, we obtain the optimal solution.
\item {\it NoReturn-ArbitraryOrigin-GivenEnd:}  In this setup, the traveler does not need to return to the origin city, and only the end city is predetermined. This problem can be solved similarly as the previous one.
\end{enumerate}

\section{Proof of Lemma~\ref{lemma:LBByBinomial}}\label{A:LBByBinomial}
Lemma~\ref{lemma:LBByBinomial} can be shown by induction. We start by considering the special case with $N=1$. In this case, by definition, we have
\begin{align}
F_X(x)=\begin{cases}
p_1, & x=1 \\
1, & x=0.
\end{cases}
\quad
F_{\hat{X}}(x)=\begin{cases}
\hat p, & x=1 \\
1, & x=0.
\end{cases}
\end{align}
Since $p_1 \geq \hat p$, the inequality $F_X(x) \geq F_{\hat X}(x)$ in \eqref{eq:FXIneq} is satisfied for $N=1$. Next, by assuming that Lemma~\ref{lemma:LBByBinomial} is true for $N=\bar{N}$, we need to show that it also holds for $N=\bar{N}+1$. For notational convenience, for $N=\bar N$, denote the ccdf of $X$ and $\hat X$ as $F_X^{\bar N}(x)$ and $F_{\hat X}^{\bar N}(x)$, respectively. Then by assumption, we have $F_X^{\bar N}(x)\geq F_{\hat X}^{\bar N}(x)$, $x=0,1,\cdots, \bar N$. As $N$ increases from $\bar N$ to $\bar N+1$, $\forall x\in \{1,\cdots, \bar N\}$, the following relationships can be obtained,
\begin{align}
F_X^{\bar N+1}(x)&=p_{\bar N+1}F_X^{\bar N}(x-1) + (1-p_{\bar N+1})F_X^{\bar N}(x), \label{eq:F1}\\
F_{\hat{X}}^{\bar N+1}(x)&=\hat{p}F_{\hat X}^{\bar N}(x-1) + (1-\hat{p})F_{\hat{X}}^{\bar N}(x). \label{eq:F2}
\end{align}
By subtracting \eqref{eq:F2} from \eqref{eq:F1} and after some manipulations, we have
\begin{align}
F_X^{\bar N+1}&(x)-F_{\hat{X}}^{\bar N+1}(x)=\left(1-\hat p\right) \left(F_X^{\bar N}(x) -F_{\hat{X}}^{\bar N}(x) \right) \notag \\
&+\hat p \left(F_X^{\bar N}(x-1) -F_{\hat{X}}^{\bar N}(x-1) \right) \notag \\
&+\left( p_{\bar N+1}-\hat {p} \right) \left(F_X^{\bar N}(x-1) -F_X^{\bar N}(x)\right) \notag \\
& \geq 0. \label{eq:LastIneq}
\end{align}
Note that the inequality in \eqref{eq:LastIneq} holds since $F_X^{\bar N}(x)\geq F_{\hat X}^{\bar N}(x)$, $p_{\bar N+1}\geq \hat p$, and $F_X^{\bar N}(x-1) \geq F_X^{\bar N}(x)$. Thus,
$\forall x\in \{1,\cdots, \bar N\}$, the inequality $F_X^{\bar N+1}(x)\geq F_{\hat{X}}^{\bar N+1}(x)$ holds. For $x=0$ or $x=\bar N+1$, the same result can be shown similarly.
This completes the proof of Lemma~\ref{lemma:LBByBinomial}.

\section{Proof of Theorem~\ref{theorem:LineSegments}}\label{A:LineSegments}
Theorem~\ref{theorem:LineSegments} can be shown by construction. Specifically, suppose that $(\mathbf q^\star(t), T^\star)$ is the optimal solution to $\mathrm{(P3)}$, and the trajectory $\mathbf q^\star(t)$ contains at least one curved segment. Then we show that there always exists an alternative solution $(\hat {\mathbf q}(t), \hat T)$ to $\mathrm {(P3)}$ such that $\hat{\mathbf q}(t)$ contains only line segments and $\hat T \leq T^\star$, as follows.

For any given optimal UAV trajectory $\mathbf q^\star(t)$, $0\leq t \leq T^\star$, define
\begin{align}
\mathcal K(t)\triangleq \{k\in \mathcal {K} : \| \mathbf q^\star(t) - \mathbf w_k\| \leq D \}.
\end{align}
In other words, for any time $t\in [0, T^\star]$, $\mathcal K(t)\subset \mathcal K$ denotes the subset of the $K$ GTs that are in connection with the UAV at time $t$, given the optimal UAV trajectory $\mathbf q^\star(t)$. Since the total number of subsets of $\mathcal K$ is $2^K$ (including the empty set), $\mathcal K(t)$ can be regarded as a time-dependent function with $2^K$ discrete values.

Let $t_1, t_2, \cdots, t_L \in (0, T^\star)$ be the $L$ critical time instances when the subset of connecting GTs changes, i.e., $t_l$ is the time instance such that $\mathcal K(t_l-\epsilon)\neq \mathcal K(t_l)$ with any arbitrarily small $\epsilon$. Then the optimal UAV trajectory $\mathbf q^\star(t)$ can be partitioned into $L+1$ portions, with the subset of connecting GTs remaining unchanged within each portion. Specifically, the $l$th portion constitutes the time interval $t\in [t_{l-1}, t_l]$ with total duration $T_l\triangleq t_l-t_{l-1}$, $l=1,\cdots, L+1$. We thus have $T^\star = \sum_{l=1}^{L+1}T_l$. For the $l$th portion of the UAV trajectory, let
\begin{align}
\mathcal K(t) & = \mathcal K_l, \ t_{l-1} \leq t \leq t_{l},\\
\hat{\mathbf q}_{l-1} & = \mathbf q^\star (t_{l-1}), \ \hat{\mathbf q}_{l}= \mathbf q^\star (t_{l}).
\end{align}
Then we show in the following that without loss of optimality to $\mathrm{(P3)}$, each of the $l$th portion of the UAV trajectory $\mathbf q^\star(t)$, $t_{l-1} \leq t \leq t_{l}$, can be replaced by the line segment connecting $\hat{\mathbf q}_{l-1}$ and $\hat{\mathbf q}_{l}$. We show this by addressing the two different cases with $\mathcal K_l =\emptyset$ or $\mathcal K_l \neq \emptyset$, separately.

{\it Case 1: $\mathcal K_l =\emptyset$}. In this case, no GT is in connection with the UAV for the $l$th portion of the UAV trajectory. As a result, this portion does not contribute to the left hand side (LHS) of the minimum connection time constraint \eqref{eq:ConstraintLB3}. Thus, replacing this trajectory portion with a line segment from $\hat{\mathbf q}_{l-1}$ to $\hat{\mathbf q}_{l}$ does not alter the feasibility of \eqref{eq:ConstraintLB3}. Furthermore, since line segment gives the shortest distance for any two given points, it is always feasible for the UAV to travel along this new segment within the time duration $\hat T_l \leq T_l$ while satisfying the maximum speed constraint \eqref{eq:speedConstr}. Thus, such a replacement ensures the feasibility of $\mathrm{(P3)}$ and at least achieves the same minimum mission completion time as $T^\star$.

{\it Case 2: $\mathcal K_l \neq \emptyset$}. In this case, those GTs in $\mathcal K_l$ are in connection with the UAV, i.e., the $l$th portion of the UAV trajectory contributes to the LHS of \eqref{eq:ConstraintLB3} for those GTs in $\mathcal K_l$. Define $\mathcal Q_l \triangleq \{\mathbf q\in \mathbb{R}^{2\times 1}: \| \mathbf q- \mathbf w_k \|\leq D, \ \forall k \in \mathcal K_l \}$, i.e., $\mathcal Q_l$ denotes the set of all possible UAV locations ensuring that all the GTs in $\mathcal K_l$ are in connection with the UAV. Note that $\mathcal Q_l$ is the intersection of $|\mathcal K_l|$ convex sets, and hence is also convex \cite{202}. As a result, since both $\hat{\mathbf q}_{l-1}$ and $\hat{\mathbf q}_{l}$ belong to the convex set $\mathcal Q_l$, then any point on the line segment connecting $\hat{\mathbf q}_{l-1}$ and $\hat{\mathbf q}_{l}$ must also belong to $\mathcal Q_l$. In other words, by replacing the original curved trajectory portion $\mathbf q^\star(t)$, $t\in [t_{l-1}, t_{l+1}]$, with the line segment connecting $\hat{\mathbf q}_{l-1}$ and $\hat{\mathbf q}_{l}$, the subset of connecting GTs $\mathcal K_l$ remains unchanged, while the UAV needs to travel a shorter distance for this portion. Thus, such a replacement ensures the feasibility of $\mathrm{(P3)}$ and at least achieves the same minimum mission completion time as $T^{\star}$.

In summary, for any given optimal solution $(\mathbf q^\star(t), T^\star)$ to $\mathrm{(P3)}$ with curved UAV trajectory, we can always construct an alterative optimal trajectory to $\mathrm{(P3)}$ by sequentially connecting the critical locations $\hat{\mathbf q}_0, \hat{\mathbf q}_{1}, \cdots, \hat{\mathbf q}_{L+1}$ with line segments, which achieves at least the same minimum mission completion time as $T^\star$.  This thus completes the proof of Theorem~\ref{theorem:LineSegments}.

\bibliographystyle{IEEEtran}
\bibliography{IEEEabrv,IEEEfull}

\end{document}